\documentclass{article}
\usepackage{graphics}
 \usepackage{graphicx}

 \usepackage{epsfig}

\usepackage{amssymb}
\usepackage{amsmath}
 \usepackage{amsthm}

\begin{document}

\title{Time-Dependent Thermopower Effect in an Interacting
Quantum Dot}

\author{M.˜ Bagheri Tagani,
        H.˜ Rahimpour Soleimani,\\
        \small{Department of physics, University of Guilan, P.O.Box 41335-1914, Rasht, Iran}}

\maketitle

\begin{abstract}
The time-dependent thermopower is analyzed through an interacting
quantum dot coupled to a time-dependent gate voltage and under the
influence of an external magnetic field using the Keldysh
nonequilibrium Green's function formalism. Formal expressions of
the electrical and  thermal conductances, thermopower, and
thermoelectrical figure of merit are obtained. The influence of
the magnetic field on the displacement current and the heat
current is studied. Results show that although applying
time-dependent gate voltage results in the enhancement of the
Seebeck coefficient, the electron-electron interaction gives rise
to a significant reduction in the thermopower. The reason for why
applying time dependent gate voltage results in the enhancement
of the thermopower is also analyzed.
\end{abstract}

\section{Introduction}
\label{Introduction} Thermopower - the ratio of induced voltage
 to an applied temperature gradient across the sample at
the state of vanishing current - is one of the oldest issues of
solid-state physics. The study of thermopower in nanoscale devices
has
 attracted a lot of attention in recent years because of
recent developments in fabricating and utilizing  them. In
addition, deviations from the Wiedmann-Franz law in nanostructures
result in significant enhancements in the thermopower  of such
devices~\cite{ref1}. Various models have been suggested for the
study of thermoelectric transport through
 quantum dots (QDs) and molecular junctions
~\cite{Dubi1,Beenakker,Galperin,Koch,Dubi2,Hsu}, nanotubes and
nanowires~\cite{Rego,Wang,Dubi3}, strongly correlated
nanostructures~\cite{Freericks,Krawiec,Peterson},  etc. In
addition,  measurement of the thermopower through nanostructures
has been an interesting topic in recent
years~\cite{Uchida,Jaworski,Zeng,Pernstich,Tan}.
\par Recently, Crepieux and co-workers~\cite{Crepieux} proposed
applying time-dependent gate voltage  results in the enhancement
of the thermopower. Their results showed that the thermopower is
enhanced by up to $40$ $\%$. Time-dependent transport has been
extensively studied both theoretically and experimentally
~\cite{Jaho,Suny,Souza,Platero,Souza1,Perfetto,Meyer,Lai}.
 However, the time-dependent heat current and
the thermopower are  novel phenomena requiring more attention. In
this paper, we consider an interacting QD coupled to  metal leads.
Although the system  seems to be similar to what was considered in
Ref.~\cite{Crepieux} , the effects of electron-electron
interactions and the magnetic field are also taken into account.
Using the Keldysh nonequilibrium Green's function
formalism~\cite{Keldysh,Jaho1}, expressions for the electrical
and  thermal conductance  are obtained. The results show that the
time-dependent thermopower is significantly reduced by
electron-electron interactions. The influence of the magnetic
field on the heat current is also investigated.
\par In the next section, the heat current is evaluated using the
nonequilibrium Green's function formalism. We use the Hartree
approximation and hence the heat current is related to the
averaged electron density. In sect. 3, numerical results are
presented and in the end,   conclusion is presented.

  \section{Model}
 \label{Model}
We consider an interacting QD coupled to metal leads and under
the influence of a step-like gate voltage pulse. The Hamiltonian
describing the system is given as follows:
\begin{equation}\label{Eq.1}
  H=\sum_{\alpha k\sigma}\varepsilon_{\alpha k\sigma} c^{\dag}_{\alpha
  k\sigma}c_{\alpha
  k\sigma}+\sum_{\sigma}\varepsilon_{\sigma}(t)n_{\sigma}+Un_{\uparrow}n_{\downarrow}+\sum_{\alpha
  k\sigma}[V_{\alpha k\sigma}c^{\dag}_{\alpha
  k\sigma}d_{\sigma}+H.C]
\end{equation}
where $c_{\alpha k\sigma} (c^{\dag}_{\alpha k\sigma})$ destroys
(creates) an electron with wave vector $k$, spin $\sigma$, and
energy $\varepsilon_{\alpha k\sigma}$ in lead $\alpha$ ( $\alpha=$
L or R ). $d_{\sigma} (d^{\dag}_{\sigma})$ is the annihilation
(creation) operator for the dot and
$n_{\sigma}=d^{\dag}_{\sigma}d_{\sigma}$ is the occupation
number. $\varepsilon_{\sigma}(t)$ denotes the time-dependent
energy level of the QD defined as
$\varepsilon_{\sigma}(t)=\varepsilon_{\sigma}^{0}+\Delta_{d}\Theta(t)$
where $\Delta_d$ is the time-variation of the gate voltage.
$\varepsilon_{\sigma}^{0}=\varepsilon_0\pm E_z$ (plus sign for
spin-up) is the time-independent energy level of the QD and $E_z$
is Zeeman splitting induced by an external magnetic field.  $U$
and $V_{\alpha k\sigma}$ stand for Coulomb repulsion and
tunneling strength between the dot and the lead $\alpha$,
respectively. The time-dependent heat current ($I^\textrm{h}$) is
obtained from the difference between the energy current
($I^\textrm{e}$) and the charge current ($I^\textrm{q}$)
according to
\begin{equation}\label{Eq.2}
  I^\textrm{h}_{\alpha}(t)=I_{\alpha}^\textrm{e}(t)-\frac{\mu_{\alpha}(t)}{e}I_{\alpha}^\textrm{q}(t)
\end{equation}
where $I^\textrm{q}_{\alpha}=-\sum_{k\sigma\in
\alpha}<\frac{\textrm{d}}{\textrm{dt}}c^{\dag}_{\alpha
k\sigma}c_{\alpha k\sigma}>$,
$I^\textrm{e}_{\alpha}=-\sum_{k\sigma\in
\alpha}\varepsilon_{\alpha
k\sigma}<\frac{\textrm{d}}{\textrm{dt}}c^{\dag}_{\alpha
k\sigma}c_{\alpha k\sigma}>$, and $\mu_{\alpha}$ denotes the
chemical potential of  lead $\alpha$. The Keldysh nonequilibrium
Green's function formalism is used to obtain the energy current.
It is straightforward to show that by means of the Green's
function of the isolated leads, the energy current is given as
($\hbar=1$)~\cite{Crepieux}
\begin{align}\label{Eq.3}
  I^\textrm{e}_{\alpha}(t)&=2Re\sum_{k\sigma\in \alpha}|\textrm{V}_{\alpha
  k\sigma}|^2\int^{t}_{-\infty}\textrm{d}t_1 \textrm{i}\varepsilon_{\alpha k\sigma}\textrm{e}^{-\textrm{i}\varepsilon_{\alpha
  k\sigma}(t_1-t)}\\ \nonumber
  &[f_{\alpha}(\varepsilon_{\alpha k\sigma})G^\textrm{r}_{\sigma\sigma}(t,t_1)+G^{<}_{\sigma\sigma}(t,t_1)]
\end{align}
where
$f_\alpha(\varepsilon)=[1+exp((\varepsilon-\mu_\alpha)/kT_\alpha)]^{-1}$
is the Fermi distribution function of the $\alpha^{th}$ lead and
$T_{\alpha}$ denotes the temperature of the lead.
$G^{r}_{\sigma\sigma'}(t,t')$ and $G^{<}_{\sigma\sigma'}(t,t')$
are the retarded and the lesser Green's functions of the
interacting QD, respectively.
\par The Green's function of the  QD
is obtained from a Dayson-like equation as
\begin{equation}\label{Eq.4}
  G_{\sigma\sigma'}(\tau,\tau')=g_{\sigma\sigma'}(\tau,\tau')+\int_{C}\textrm{d}t_1\textrm{d}t_2g_{\sigma\sigma'}(\tau,\tau_1)
  \Sigma_{\sigma}(\tau_1,\tau_2)G_{\sigma\sigma'}(\tau_2,\tau)
\end{equation}
where $C$ denotes the contour integral, and
$\Sigma_{\sigma}(t_1,t_2)=\sum_{\alpha k}|V_{\alpha
k\sigma}|^{2}g_{\alpha k\sigma}(t_1,t_2)$ is the self-energy
operator in which $g_{\alpha k\sigma}(t,t')$ is the isolated lead
Green's function. $g_{\sigma\sigma'}$ is the isolated QD Green's
function satisfying ~\cite{explain1}
\begin{equation}\label{Eq.5}
  [\textrm{i}\frac{\textrm{d}}{\textrm{d}t}-\varepsilon_{\sigma}(t)-U<n_{\bar{\sigma}}(t)>]g_{\sigma\sigma'}(t,t')=\delta(t-t')\delta_{\sigma\sigma'}
\end{equation}
For deriving the above equation, the decoupling approximation
introduced in Ref.~\cite{Suny} has been used, i.e.,
$<\{n_{\bar{\sigma}}(t)d_{\sigma}(t),d^{\dag}_{\sigma'}(t')\}>\approx<n_{\bar{\sigma}}(t)>g_{\sigma\sigma'}(t,t')$.
 $<n_{\sigma}(t)>$ denotes the time averaged electron
density with  spin $\sigma$, and $\bar{\sigma}$ is opposite of
$\sigma$. This approximation is reasonable under conditions that
the temperature is low enough (lower than the level spacing) and
the bias voltage is small. In the following, we change the sum
over $k$ into an energy integral and use the wide band
approximation, i.e., $\sum_{k}|V_{\alpha k\sigma}|^2=\int
d\varepsilon/2\pi \Gamma_{\alpha}^{\sigma}$ where
$\Gamma_{\alpha}^{\sigma}=2\pi\rho_{\alpha}|V_{\alpha
k\sigma}|^2$ is the spin-dependent tunneling rate. The retarded
and  lesser Green functions are obtained from  Eq. 4 by means of
the Langreth continuation theorem~\cite{Langreth}. Then, the
time-dependent heat current is given as
\begin{align}\label{Eq.6}
  I^\textrm{h}_{\alpha}(t)&=-\sum_{\sigma}\Gamma_{\alpha}^{\sigma}[\sum_{\alpha'}\Gamma_{\alpha'}^{\sigma}\int\frac{\textrm{d}\varepsilon}{2\pi}(\varepsilon-\mu_{\alpha})f_{\alpha'}(\varepsilon)|A_{\alpha'}^{\sigma}(\varepsilon,t)|^2\\
  \nonumber
  &+\int\frac{\textrm{d}\varepsilon}{\pi}(\varepsilon-\mu_{\alpha})f_{\alpha}(\varepsilon)\textrm{Im}\{A_{\alpha}^{\sigma}(\varepsilon,t)\}]
\end{align}
where $A_{\alpha}^{\sigma}(\varepsilon,t)$ is given by~\cite{Jaho}
\begin{equation}\label{Eq.7}
  A_{\alpha}^{\sigma}(\varepsilon,t)=\frac{[\varepsilon-\varepsilon_{n\sigma}+\textrm{i}/2\Gamma^{\sigma}]-\Delta_d\textrm{e}^{\textrm{i}[\varepsilon-\varepsilon_{n\sigma}-\Delta_d+\textrm{i}/2\Gamma^{\sigma}]t}}
  {[\varepsilon-\varepsilon_{n\sigma}+\textrm{i}/2\Gamma^{\sigma}][\varepsilon-\varepsilon_{n\sigma}-\Delta_d+\textrm{i}/2\Gamma^{\sigma}]}
\end{equation}
where
$\varepsilon_{n\sigma}=\varepsilon_{\sigma}+U<n_{\bar{\sigma}}>$
and $\Gamma^{\sigma}=\Gamma_{L}^{\sigma}+\Gamma^{\sigma}_{R}$. It
is straightforward to show the time averaged  electron density is
obtained from~\cite{Jaho}
\begin{align}\label{new}
  <n_{\sigma}(t)>&=\lim_{T\rightarrow\infty}\frac{1}{2T}\sum_{\alpha}\Gamma^{\sigma}_{\alpha}\int\frac{\textrm{d}\varepsilon}{2\pi}
  f_{\alpha}(\varepsilon)\int_{-T}^{T}\textrm{d}t|A_{\alpha}^\sigma(\varepsilon,t)|^2\\\nonumber
  &=\sum_{\alpha}\int\frac{\textrm{d}\varepsilon}{2\pi}f_{\alpha}(\varepsilon)\frac{(\varepsilon-\varepsilon_{n\sigma})^2+\Delta_d^2+(\frac{\Gamma^{\sigma}}{2})^2}
  {[(\varepsilon-\varepsilon_{n\sigma})^2+(\frac{\Gamma^{\sigma}}{2})^2][(\varepsilon-\varepsilon_{n\sigma}-\Delta_d)^2+(\frac{\Gamma^{\sigma}}{2})^2]}
\end{align}
Equation 8 should be solved self-consistently to obtain the
 electron density. In the following, we assume that
$\Gamma_{\alpha}^{\sigma}=\Gamma_0$ and use $\Gamma_0$ as the
energy unit~\cite{explain2} and $\hbar/\Gamma_0$ as the time unit.
\begin{figure}
\begin{center}
\includegraphics[height=100mm,width=120mm,angle=0]{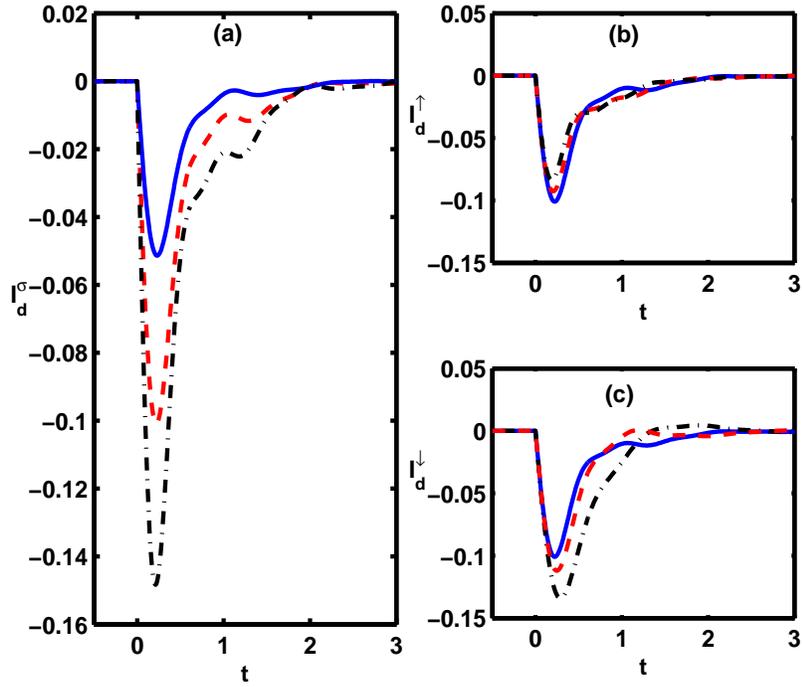}
\caption{(a) $I_d^{\sigma}$ versus time for $\Delta_d=0.5$
(solid), $\Delta_d=1$ (dashed) and $\Delta_d=1.5$ (dashed-dotted).
(b) and (c) show spin-resolved displacement current for $E_z=0$
(solid), $E_z=1$ (dashed) and $E_z=2$ (dashed-dotted). Parameters
are $T_1=2, T_2=0.1,\varepsilon_0=0, U=2$, and
$\mu_{L(R)}=+(-)2.5$. For (b) and (c), we set $\Delta_d=1$.}
\label{fig:1}
\end{center}      
\end{figure}
\section{Results and Discussion }
\label{Numerical results}

Figure 1 shows the spin-dependent displacement current
($I_\textrm{d}^{\sigma}(t)=I_\textrm{L}^{\sigma}(t)+I_\textrm{R}^{\sigma}(t)$)
as a function of the amplitude of the gate voltage ($\Delta_d$)
(Fig. 1a) or the magnetic field (Figs. 1b and c). The
displacement current describes the time evolution of the electron
density of the QD. As we expect, $I_\textrm{d}$ is zero at $t<0$,
because the system is in the steady state. Upon applying the gate
voltage , a significant reduction in $I_\textrm{d}^\sigma$ is
observed when $0<t<2$. The gate voltage shifts the  QD's energy
level toward the chemical potential of the emitter (left lead)
and hence, a sudden decrease in the population of the charge
occurs. The reduction of the population leads to
$I_\textrm{d}^\sigma<0$. During the time, the system approaches
the new steady state and, therefore, $I_\textrm{d}^\sigma$
becomes zero again. This behavior of the displacement current was
recently reported in Ref.~\cite{Lai}. Unlike the results obtained
in Ref.~\cite{Cuansing}, the displacement current does not show
any fluctuation. Indeed, in Ref.~\cite{Cuansing} the coupling at
$t=0$ adds some energy to the system but here, the gate voltage
just changes the position of energy levels. Figures 1b and c
describe the behavior of the spin-up and spin-down displacement
currents in the response to the external magnetic field,
respectively. It is observed that $I_\textrm{d}^\uparrow$ is not
sensitive to the magnetic field. Indeed, in the presence of the
magnetic field, the population of the spin-up is significantly
reduced, because the entrance of the spin-up electron into the QD
requires  the energy to be more than $\mu_\textrm{L}$ due to
electron-electron interactions. The behavior of
$I_\textrm{d}^\downarrow$ is more interesting in the response to
the magnetic field. Unlike the spin-up level, the spin-down level
is always inside the bias window and the  increasing magnetic
field gives rise to the enhancement of the spin-down electron
population. It is found that $I_\textrm{d}^\downarrow$ becomes
positive in a strong magnetic field ($E_z=2$) when $1.5<t<2.5$. It
comes from  the inequality in the temperature of the leads.
\begin{figure}[htb]
\begin{center}
\includegraphics[height=90mm,width=120mm,angle=0]{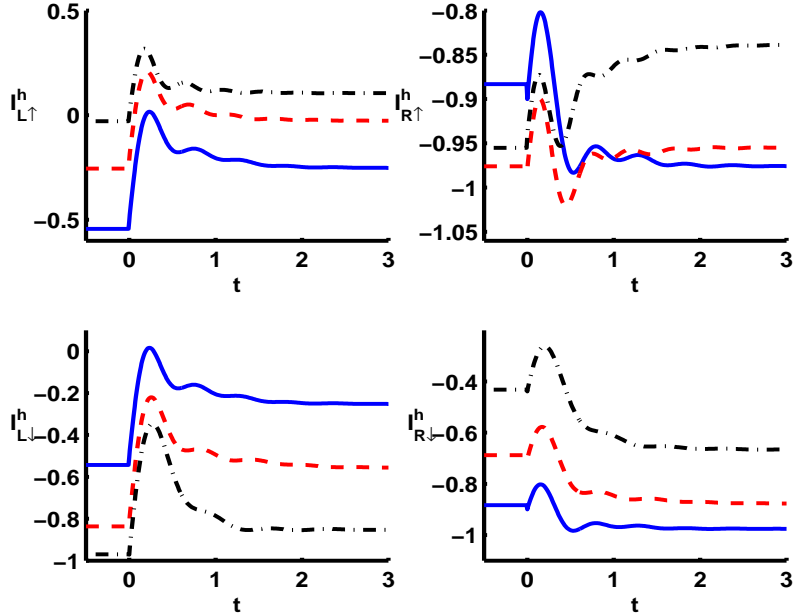}
\caption{Spin-dependent heat current against time. $E_z=0$
(solid), $E_z=1$ (dashed) and $E_z=2$ (dashed-dotted). Other
parameters are the same as Fig. 1.}
\end{center}
\end{figure}
\par The spin resolved heat current is plotted in  Fig. 2 for
different magnetic fields.  The typical coherent oscillations in
the heat current are observed. It is interesting to note that the
frequency of the oscillations is spin- and Coulomb
repulsion-dependent and given by
$\hbar\omega_{\sigma}=|E_f-\varepsilon_{\sigma}-U<n_{\bar{\sigma}}>-\Delta_\textrm{d}|$.
The existence of such beats in the charge current was previously
reported~\cite{Souza,Lai}; however, for the heat current more
experimental results are needed. At $t<0$, or $t\rightarrow
\infty$, the heat current approaches  constant values. It is also
observed that the variations of the heat current in the right
lead are more significant than in the left one because of
$T_\textrm{R}<T_\textrm{L}$. Notice, the heat current becomes
positive when the energy level of the QD is outside the bias
window. Therefore, $I^\textrm{h}_{\textrm{L}\uparrow}>0$ for
$E_z=1$ and $2$. In the case of
$I^\textrm{h}_{\textrm{L}\downarrow}$, the magnitude of the heat
current is enhanced by an increase of the magnetic field because
of $\mu_\textrm{L}>>\varepsilon_{\downarrow}$. The story is
completely different about $I^\textrm{h}_{\textrm{R}\sigma}$,
because an increase of magnetic field leads to
$\varepsilon_\sigma>\mu_\textrm{R}$ and hence, the magnitude of
the heat current is decreased.
\par In the linear response limit, the current is given by $I(t)=G_V(t)\Delta V+G_T(t)\Delta
T$ where $G_V(t)$ and $G_T(t)$ are the electrical conductance and
the thermal coefficient, respectively. Setting $I(t)=0$, the
Seebeck coefficient is defined as
\begin{equation}\label{Eq.8}
  S(t)=-\frac{\Delta V}{\Delta T}=\frac{G_T(t)}{G_V(t)}
\end{equation}
From  Eq. 6, we obtain the time-dependent conductance coefficients
as follows~\cite{explain}:
\begin{subequations}\label{Eq.9}
\begin{align}
G_V(t)&=-\sum_{\sigma}\frac{\Gamma_\textrm{L}^{\sigma}\Gamma^{\sigma}_{\textrm{R}}}{\Gamma^{\sigma}}\int
\frac{\textrm{d}\varepsilon}{\pi}f'(\varepsilon)\textrm{Im}(A^\sigma(\varepsilon,t))
\\
G_T(t)&=\frac{1}{T}\sum_{\sigma}\frac{\Gamma_\textrm{L}^{\sigma}\Gamma^{\sigma}_{\textrm{R}}}{\Gamma^{\sigma}}\int
\frac{\textrm{d}\varepsilon}{\pi}(\varepsilon-\varepsilon_F)f'(\varepsilon)\textrm{Im}(A^\sigma(\varepsilon,t))
\end{align}
\end{subequations}
\begin{figure}[htb]
\begin{center}
\includegraphics[height=90mm,width=120mm,angle=0]{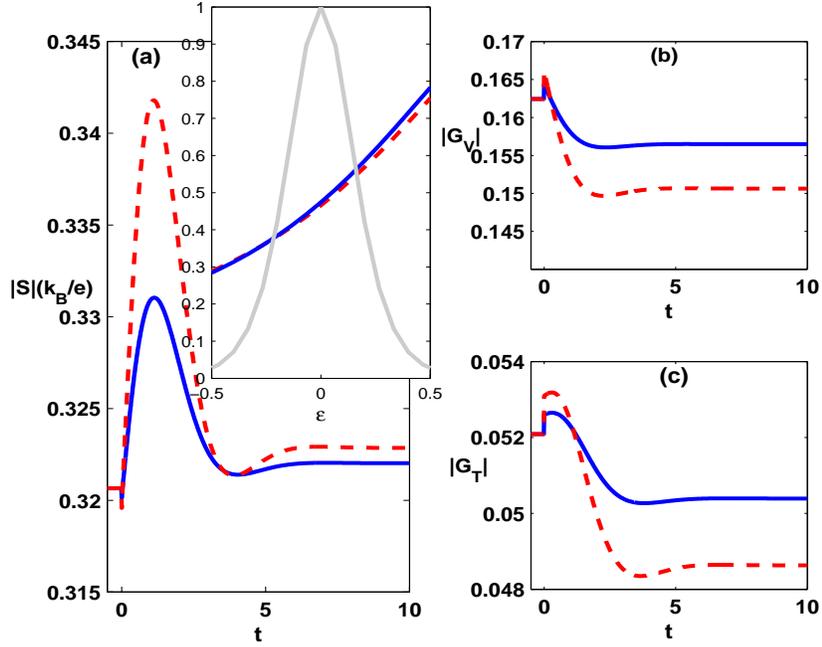}
\caption{a) Thermopower, (b) conductance, and (c) thermal
coefficient versus time. Parameters are $T_1=T_2=0.1,
\varepsilon_0=0.5, E_z=0, U=2$ and $\Delta_\textrm{d}=0.05$
(solid) and $\Delta_\textrm{d}=0.1$ (dashed). Inset shows the
generalized spectral function as  a function of energy at $t=1$
(solid line) and $t=7$ (dashed line). Normalized
$f'(\varepsilon)$ is also plotted in gray.}
\end{center}
\end{figure}
where $f'(\varepsilon)=\textrm{d}/\textrm{d}\varepsilon
f(\varepsilon)$. Note that the above equations are only valid
under conditions that $\Delta V$ and $\Delta T$ are small in
comparison to $\varepsilon_0$ and $\Delta_d^2\rightarrow 0$. In
these conditions, we have
$|A^{\sigma}(\varepsilon,t)|^2=-2/\Gamma^{\sigma}\textrm{Im}[A^{\sigma}(\varepsilon,t)]$~\cite{Crepieux}.
It is interesting to note that the imaginary part of
$A(\varepsilon,t)$ is thought of as the generalized spectral
function. The time-dependent Seebeck coefficient is plotted in
Fig. 3a. One observes that there is a sudden increase in
thermopower once the gate voltage  is applied. Such a
time-dependent enhancement has been recently reported in a
non-interacting QD~\cite{Crepieux}. Indeed, the more sensitivity
of the system to the temperature in the transient regime results
in the enhancement of the thermopower. Since $\Delta_\textrm{d}$
is small, the imaginary part of $A^\sigma(\varepsilon,t)$ is
composed of a dominant Lorentzian-like part where
$\varepsilon_{n\sigma}$ is its center, and a corrective part
which is on the order of $\Delta_\textrm{d}$ and decays during the
time according to $\textrm{e}^{-1/2\Gamma^{\sigma}t}$. It is
interesting to note that the correction part plays the main role
in the enhancement of the thermopower. $f'(\varepsilon)$ has a
Lorentzian shape which is centered near the chemical potential of
the lead, see inset of Fig. 3. Under these conditions, the
corrective term causes  $\textrm{Im}(A^{\sigma}(\varepsilon,t))$
to become larger at initial times after  applying gate voltage.
It leads to the enhancement of the thermopower. This term decays
with time and, as a result, the thermopower becomes constant
again. Indeed, the increasing generalized spectral function near
the chemical potential of the leads results in the enhancement of
the thermopower upon  applying gate voltage. The generalized
spectral function and $f'(\varepsilon)$ are plotted in the inset
of Fig. 3a. It is evident that the increase of $\Delta_\textrm{d}$
leads to  more increase of the generalized spectral function and,
as a result, the thermopower is more enhanced.

 Figs. 3b and c show  the time
evolution of $G_V$ and $G_T$ as a function of $\Delta_\textrm{d}$,
respectively. It is observed that an increase in
$\Delta_\textrm{d}$ leads to an increase of the variation
amplitude of $G_V$ and $G_T$.

\par The influence of the Coulomb interaction on the thermopower is analyzed
in  Fig. 4 as a percentage of
$(S^\textrm{max}-S^\textrm{sat})/S^\textrm{sat}$ where
$S^\textrm{max}$ is the maximum value of the thermopower and
$S^\textrm{sat}=S(t\rightarrow\infty)$. It is found that the
thermopower is reduced by up to $25$ $\%$ for strong
electron-electron interactions. This reduction is more significant
at high $\Delta_d$. Therefore, the predicted enhancement of the
thermopower up to $40$ $\%$  cannot  be observed in  strong
electron-electron interactions. Indeed, an increase of correlation
between electrons results in a decrease of the thermopower.
\begin{figure}[htb]
\begin{center}
\includegraphics[height=80mm,width=95mm,angle=0]{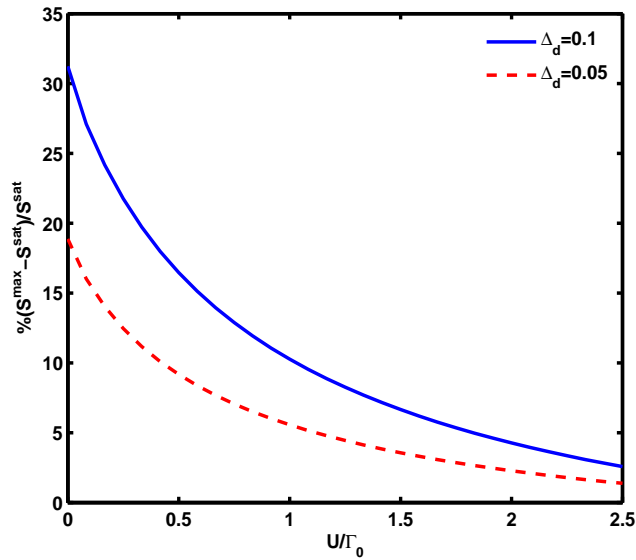}
\caption{Percentage of
$(S^\textrm{max}-S(t\rightarrow\infty))/S(t\rightarrow\infty)$.
We set $\varepsilon_0=0.2$. Other parameters are the same as in
Fig. 3.}
\end{center}
\end{figure}

\section{Conclusion}
\label{conclusion} In this paper, we analyze  the time-dependent
Seebeck coefficient through an interacting quantum dot subject to
a magnetic field. The formal expression of the thermopower is
obtained using the nonequilibrium Green's function formalism. The
influence of the magnetic field on the displacement and heat
currents is examined, and it is observed that the current of the
left and  right leads are different from each other in the
response to the magnetic field. Spin-dependent beats in the heat
current are also observed. We find that the thermopower is
reduced by up to $25$ $\%$ under certain conditions. The behavior
of the electrical and  thermal conductances in the response to the
time-dependent gate voltage are also examined.


\begin{thebibliography}{}

\bibitem{ref1}
B. Kubala, J. K\"{o¨}nig,  J. Pekola,  Phys. Rev. Lett.
\textbf{100}, 066801 (2008)


\bibitem{Dubi1}
Y. Dubi,  M. Di Ventra,  Rev. Mod. Phys. \textbf{83}, 131 (2011)

\bibitem{Beenakker}
C. W. J. Beenakker, A. A. M. Starling,  Phys. Rev. B \textbf{46},
9667–9676 (1992)

\bibitem{Galperin}
M. Galperin, A. Nitzan, M. A. Ratner, Phys. Rev. B \textbf{75},
155312 (2007)

\bibitem{Koch}
J. Koch, F. von Oppen, Y. Oreg, E. Sela,  Phys. Rev. B
\textbf{70}, 195107 (2004)

\bibitem{Dubi2}
Y. Dubi, M. Di Ventra,  Phys. Rev. B  \textbf{79}, 081302 (2009)

\bibitem{Hsu}
B. C. Hsu, Y. S. Liu, S. H. Lin, Y. C. Chen,  Phys. Rev. B
\textbf{83}, 041404 (2011)

\bibitem{Rego}
L. G. C. Rego, G. Kirczenow,   Phys. Rev. Lett. \textbf{81}, 232
(1998)

\bibitem{Wang}
J. S. Wang,. J. Wang, N. Zeng,  Phys. Rev. B \textbf{74}, 033408
(2006)

\bibitem{Dubi3}
Y. Dubi, M. Di Ventra,  Nano Lett. \textbf{9(1)}, 97 (2009)

\bibitem{Freericks}
J. K. Freericks, V. Zlatic, A. M. Shvaika, Phys. Rev. B
\textbf{75}, 035133 (2007)

\bibitem{Krawiec}
M. Krawiec, K. I. Wysokinsk,  Phys. Rev. B \textbf{73}, 075307
(2006)

\bibitem{Peterson}
M. R. Peterson, S. Mukerjee, B. S. Shastry, J. O. Haerter, Phys.
Rev. B \textbf{76}, 125110 (2007)

\bibitem{Uchida}
K. Uchida, J. Xiao, H. Adachi, J. Ohe, S. Takahashi, J. Ieda, T.
Ota, Y. Kajiwara, H. Umezawa, H. Kawai, G. E. W. Bauer, S.
Maekawa, E. Saitoh,  Nat. Mater. \textbf{9},
 894 (2010)

\bibitem{Jaworski}
C. M. Jaworski, J. Yang, S. Mack, D. D. Awschalom, J. P.
Heremans, R. C. Myers, Nat. Mater.  \textbf{9}, 898 (2010)

\bibitem{Zeng}
G. Zeng, J. M. O. Zide, W. Kim, J. E. Bowers, A. C. Gossard, Z.
Bian, Y. Zhang, A. Shakouri, S. L. Singer,   A. Majumdar,  J.
Appl. Phys. \textbf{101}, 034502 (2007)

\bibitem{Pernstich}
K. P. Pernstich, B. R$\ddot{o}$ssner, B. Batlogg,
 Nat. mater. \textbf{7}, 321 (2008)

\bibitem{Tan}
A. Tan, S. Sadat, P. Reddy,  Appl. Phys. Lett. \textbf{96}, 013110
(2010)

\bibitem{Crepieux}
A. Cr$\acute{e}$pieux, F. Simkovic, B. Cambon,  F. Michelini,
 Phys.
Rev. B \textbf{83}, 153417 (2011)

\bibitem{Jaho}
A. P. Jauho, N. S. Wingreen, Y. Meir,  Phys. Rev. B \textbf{50},
5528 (1994)

\bibitem{Suny}
Q. F. Sun, T. H. Lin,  J. Phys.: Condens. Matter \textbf{9}, 4875
(1997)

\bibitem{Souza}
F. M. Souza, Phys. Rev. B \textbf{76}, 205315 (2007)

\bibitem{Platero}
G. Platero,  R. Aguado,  Phys. Rep. \textbf{395}, 1 (2004)

\bibitem{Souza1}
F. M. Souza, S. A. Le$\tilde{a}$o, R. M. Gester,  A. P. Jauho,
 Phys. Rev. B \textbf{76}, 125318 (2007)

\bibitem{Perfetto}
E. Perfetto, G. Stefanucci,  M. Cini,  Phys. Rev. B \textbf{78},
155301 (2008)

\bibitem{Meyer}
C. Meyer, J. M. Elzerman, L. P. Kouwenhoven,  Nano Lett.
\textbf{7}, 295 (2007)

\bibitem{Lai}
W. T. Lai, D. M. T. Kuo, P. W Li,  Physica E \textbf{41}, 886
(2009)

\bibitem{Keldysh}
L. V. Keldysh,  Zh. Eksp. Teor. Fiz. \textbf{47}, 1515 (1964)

\bibitem{Jaho1}
H. Haug, A.P. Jauho, \textit{Quantum Kinetics in Transport and
Optics of Semiconductors} (Springer, Heidelberg, 1996)

\bibitem{explain1}
For deriving  Eq. 5, We have used decoupling approximation
introduced in Ref.~\cite{Suny}.

\bibitem{Langreth}
D. C. Langreth, \textit{in Linear and Nonlinear Electron
Transport in Solids}, Vol. 17 of Nato Advanced Study Institute
Series B: Physics, ed. by J. T. Devreese, V. E. Van Doren
(Plenum, New York, 1976)

\bibitem{explain2}
For generic values of $\Gamma_0$, see for example, D. G. Gordon,
H. Shtrikman, D. Mahalu, D. A. Magder, U. Meirav, M. A. Kastner,
 Nature \textbf{391}, 156
(1998)

\bibitem{Cuansing}
E. C. Cuansing, J. S. Wang,  Phys. Rev. B \textbf{81}, 052302
(2010)

\bibitem{explain}
For deriving a relation for $G_V$ and $G_T$, we assume that
$\mu_{\textrm{L}}=E_f+\Delta V$,  $T_\textrm{L}=T+\Delta T$,
$\mu_\textrm{R}=E_f$ and $T_\textrm{R}=T$. Therefore, Fermi
functions of leads can be written as $f_\textrm{L}=f+\Delta V
f'-\Delta T/T f'$ and $f_\textrm{R}=f$.

 \end{thebibliography}
\end{document}